\documentstyle[12pt,psfig]{article}

\newcommand{\beq}{\begin{equation}}
\newcommand{\eeq}{\end{equation}}
\newcommand{\bea}{\begin{eqnarray}}
\newcommand{\eea}{\end{eqnarray}}

\newcommand{\gsim}{\lower.7ex\hbox{$
\;\stackrel{\textstyle>}{\sim}\;$}}
\newcommand{\lsim}{\lower.7ex\hbox{$
\;\stackrel{\textstyle<}{\sim}\;$}}

\def\lsim{\mathrel{\rlap{\lower3pt\hbox{\hskip0pt$\sim$}}
    \raise1pt\hbox{$<$}}}         
\def\gsim{\mathrel{\rlap{\lower4pt\hbox{\hskip1pt$\sim$}}
    \raise1pt\hbox{$>$}}}         

\newcommand{\bibit}[1]{\bibitem{#1}}

\newcommand{\aver}[1]{\langle #1\rangle}

\newcommand{\La}{\overline{\Lambda}}

\newcommand{\GeV}{\,\mbox{GeV}}
\newcommand{\MeV}{\,\mbox{MeV}}

\newcommand{\msp}[1]{\mbox{\hspace*{#1mm}~}}

\begin{document}
\thispagestyle{empty}
\vspace*{-20mm}

\begin{flushright}
Bicocca-FT-03-21\\
UND-HEP-03-BIG\hspace*{.08em}03\\
hep-ph/0308165\\
\vspace*{2mm}
\end{flushright}
\vspace*{15mm}

\boldmath
\begin{center}
{\LARGE{\bf
On Extracting Heavy Quark Parameters \vspace*{5mm}\\
from Moments with Cuts}}
\vspace*{9mm} 

{\tt 
Contributed to the Lepton-Photon Conference, August 11-15, 2003\\
FNAL, Batavia IL
}
\vspace*{3mm} 
\end{center}

\unboldmath
\smallskip
\begin{center}
{\large{I.I.~Bigi$^{\,a}$ and N.~Uraltsev$^{\,b\,*}$
}}  \\
\vspace{4mm}
$^a$ {\sl Department of Physics, University of Notre Dame du Lac}
\vspace*{-.8mm}\\
{\sl Notre Dame, IN 46556, USA}\vspace*{.5mm}\\
$^b$ {\sl INFN, Sezione di Milano, Milan, Italy} 
\vspace*{18mm}

{\bf Abstract}\vspace*{-.9mm}\\
\end{center}
\noindent
We point out that applying the photon energy cut 
significantly modifies the moments of energy spectrum in $B\!\to\!
X_s+\gamma$ decays, with a certain class of effects not accounted for
in the mostly used OPE expressions. This leads to a systematic bias
in the extracted values of the $b$ quark mass and other heavy
quark parameters. The apparent $b$ quark mass increases typically by
$70\MeV$ or more, together with an even more dramatic downward 
shift in the kinetic expectation value.
Accounting for these
cut-related shifts brings different measurements into a good
agreement, when the OPE-based theory employs the robust approach.
These nonperturbative effects are exponential in the effective hardness
severely lowered by high cuts, and do not signify a breakdown of
the $\,1/m_b$ expansion itself. 
Similar effects in semileptonic
$b\!\to\! c$ decays are briefly addressed.
We stress the utility of the second moment of $\,E_\gamma\,$ once these 
effects are incorporated.

\setcounter{page}{0}

\vfill

~\hspace*{-12.5mm}\hrulefill \hspace*{-1.2mm} \\
\footnotesize{
\hspace*{-5mm}$^*$On leave of absence from Department of 
Physics, University of Notre Dame, Notre Dame, 
IN 46556, USA \hspace*{-7mm}~\\
\hspace*{-5mm}\hspace*{.25em} and St.\,Petersburg Nuclear Physics 
Institute, Gatchina, St.\,Petersburg  188300, Russia}
\normalsize

\newpage

Treating effects of strong interactions in decays of beauty
particles is of primary importance nowadays when many precision
experimental studies are being carried out. Inclusive distributions
in radiative and semileptonic decays are the portal to accurately
determining the nonperturbative heavy quark parameters controlling
short-distance observables in $B$ decays. Their utility rests on an
consistent expansion in $1/m_b$, the inverse b quark mass 
\cite{theory}. Recent
experimental data generally show a nontrivial agreement between
quite different -- and {\it a priori}\, unrelated -- measurements at
the nonperturbative level, on one hand, and consistency with the
QCD-based OPE theory.

In order to enjoy in full the potential of a small expansion
parameter provided by the heavy quark mass, 
the observable in question must be sufficiently inclusive. However
experimental cuts imposed for practical reasons -- to suppress
backgrounds etc. -- often essentially degrade the effective hardness
${\cal Q}$ of the process.  This brings in another expansion
parameter $1/{\cal Q}$ effectively replacing $1/m_b$ in certain QCD
effects. The reliability of the expansion greatly deteriorates for
${\cal Q}\!\ll\! m_b$.  This phenomenon is particularly important in
$b\!\to\! s+\gamma$ decays where experiments so far have imposed cut
$E_\gamma\! >\!2\GeV$ or even higher.

The theoretical aspects of such limitations have been discussed
during the last couple of years \cite{uses,amst,ckm03}. In
particular, the effective `hardness' of the inclusive 
$b\to s+\gamma$ decays with $E_\gamma\! >\! 2\GeV$ amounts to only about
$1.25\GeV$, which casts doubts on the precision of the routinely used
expressions incorporated into the fits of heavy quark parameters.

In the present letter we point out that these effects are
numerically significant, lead to a systematic bias that often exceeds naive 
error estimates and therefore cannot be ignored. 
Evaluating them in the most straightforward
(although somewhat simplified) way we find, for instance for $b\to
s+\gamma$ decays 
\bea
\nonumber
\tilde m_b  & \msp{-3}\simeq \msp{-3} & m_b+ 70\MeV \\
\tilde \mu_\pi^2  & \msp{-3}\simeq \msp{-3} &  
\mu_\pi^2 - (0.15\div 0.2)\GeV^2
\label{12}
\eea
where $\tilde m_b$ and $\tilde \mu_\pi^2$ are the {\tt apparent}
values of the $b$ quark mass and of the kinetic expectation value,
respectively, as extracted from the $b\!\to\! s+\gamma$ spectrum
with $E_{\gamma} \!>\! 2\GeV$ in a usual way.  
Correcting for these effects eliminates
alleged problems for the OPE in describing different data and rather
leads to a too good agreement between the data on different types of
inclusive decays. 

Moreover, this resolves the controversy 
noted previously: while the values of $\La$ and $\mu_\pi^2$
reportedly extracted from the CLEO $b\to s+\gamma$ spectrum were
found to be significantly below the theoretical expectations, the
theoretically obtained spectrum itself turned out to yield a good 
description of the observed spectrum when we evaluated it based on
these theoretically preferred values of parameters \cite{uses}. 
 
The bias in Eqs.~(\ref{12}) depends on the position of
the cut (more precisely, on the gap $\frac{m_b}{2}\!-\!E_{\rm cut}$) and
the actual values of other heavy quark parameters. The quoted
estimates assume moderate values, $m_b(1\GeV)\!\simeq\! 4.6\GeV$ and
$\mu_\pi^2(1\GeV)\!\simeq\! 0.43\GeV^2$. If $m_b$ becomes lower and/or the 
true $\mu_\pi^2$ increases as may be indicated by the most recent
data, the bias increases further. 
\newpage

\noindent
{\bf OPE and cuts}~~
The origin of these effects and why they
are missed in the standard application of the OPE and in estimates
of the theoretical accuracy, have been discussed elsewhere
\cite{ckm03}. In brief, considering a 
constrained fraction of the $B\!\to\! X_s \!+\!\gamma $ events
\beq
1\!-\!\Phi_\gamma(E) = \frac{1}{\Gamma_{bs\gamma}}\,
\int_E^{\frac{M_B}{2}} {\rm d}E_\gamma \, \frac{{\rm 
d}\Gamma_{bs\gamma}} {{\rm d}E_{\gamma}}\;.
\label{110uses}
\eeq
(or similarly truncated photon energy moments), the simple-minded approach
routinely expands the spectrum in powers of
$1/m_b$. Ignoring perturbative bremsstrahlung one obtains  a
$\delta$-like spectrum peculiar for two-body decays, and the expansion 
around the free-quark kinematics does not change this  -- it  
only generates higher derivatives of $\;\delta(E_\gamma\!-\!\frac{m_b}{2})\,$:
\beq
\frac{1}{\,\Gamma^0_{bs\gamma}\!\!}\: \frac{{\rm d}\Gamma_{\!bs\gamma}} 
{{\rm d}E_{\gamma}} = a\,\delta(E_\gamma\!-\!\mbox{$\frac{m_b}{2}$}) +
b\,\delta'(E_\gamma\!-\!\mbox{$\frac{m_b}{2}$}) 
+ c\,
\delta''(E_\gamma\!-\!\mbox{$\frac{m_b}{2}$}) + ...
\label{112uses}
\eeq
where $a$, $b$, ... are given by the $B$ meson expectation values of local 
$b$-quark operators. Naively computing $1\!-\!\Phi_\gamma(E)$,
or spectral moments over the restricted domain in this way would
yield unity in Eq.~(\ref{110uses}) for any $E \!>\! \frac{m_b}{2}$ --
a result clearly meaningless on physical grounds.  The actual
behavior of the spectrum and the moments is described by the heavy
quark distribution function. Its tail is indeed exponentially
suppressed by a typical factor $e^{-c\, {\cal Q}/\mu_{\rm hadr}}\,$ at
$\,{\cal Q}(E_\gamma)\!\gg\! \mu_{\rm hadr\,}$. However, for 
$\,{\cal Q}(E_\gamma) \!\sim\!  \mu_{\rm hadr}\,$ there is little
suppression of the missed tail contribution; the error becomes of order one.

This obvious point is missed in the naive application of the OPE and
in the way to gauge the theoretical uncertainty based on it.
Conceptually this is related to the limited range of convergence of
the OPE for the width, determined in this case by the support of the
heavy quark distribution function \cite{ckm03}.
\vspace*{.7mm}

Here we rather concentrate on the numerical consequences for 
$B\!\to\! X_s+\gamma$ decays. To this end we first turn off perturbative
effects altogether. The spectrum then is described by the
nonperturbative light-cone distribution function $F(k_+)$:
\beq
\frac{1}{\Gamma}\frac{{\rm d}\Gamma}{{\rm d}E_\gamma} = 2F(2E_\gamma-m_b) 
\;.
\label{22}
\eeq
Although not necessary for our purpose, one can imagine a
theoretical heavy quark limit with fixed hardness ${\cal Q}\,$:
\beq
{\cal Q} \equiv m_b\!-\!2E_{\rm cut}= \mbox{fixed}\;, \msp{7} m_b\to \infty\;.
\label{24}
\eeq
The fully integrated moments of $F(x)$ and therefore of 
the spectrum give then directly the underlying heavy quark parameters:
\beq
\int_{0}^{\infty} k\, F(\La-k)\: {\rm d}k  = \La, \msp{10}
\int_{0}^{\infty} (k-\La)^2\, F(\La-k)\: {\rm d}k = \frac{\mu_\pi^2}{3},
\mbox{ etc.}
\label{26}
\eeq
As mentioned above, in the standard practical-OPE--based formulae 
these relations remain the same for the moments 
with the cut as well
(provided $E_{\rm cut}\!<\! \frac{m_b}{2}$ which is always assumed) --
yet not in reality. Paralleling the routinely used way we therefore
introduce
\beq
\tilde\Lambda(E_{\rm cut})\!=\! \frac{\int_{E_{\rm cut}} 
\mbox{{\small$(M_B\!-\!2E_\gamma)\,$}}
\frac{{\rm d}\Gamma}{{\rm d}E_\gamma}\:{\rm d}E_\gamma}{\int_{E_{\rm cut}}
\frac{{\rm d}\Gamma}{{\rm d}E_\gamma}\:{\rm d}E_\gamma} ,\msp{4.2} 
\tilde\mu_\pi^2 (E_{\rm cut})\!=\! 3\!\left[
\frac{\int_{E_{\rm cut}} \mbox{{\small$(M_B\!-\!2E_\gamma)^2$}}
\frac{{\rm d}\Gamma}{{\rm d}E_\gamma}\:{\rm d}E_\gamma}{\int_{E_{\rm cut}}
\frac{{\rm d}\Gamma}{{\rm d}E_\gamma}\:{\rm d}E_\gamma}
\!-\!\tilde\Lambda^2(E_{\rm cut}) \right]\!.
\label{28}
\eeq
Clearly the {\tt apparent} value $\tilde\Lambda(E_{\rm cut})$ is
always below the actual $\La$. This is illustrated by Fig.~1a where
we plot the cut-related `bias' -- the difference between $\La$ and 
$\tilde\Lambda(E_{\rm cut})$ as a function of energy $E_{\rm cut}$. It 
turns out to be quite significant.

\begin{figure}[hhh]
\begin{center}
\mbox{\psfig{file=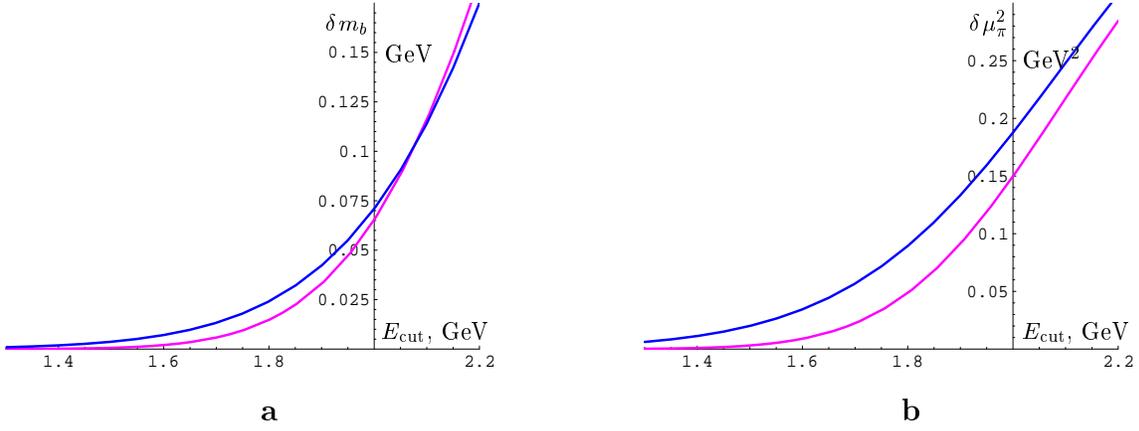,width=150mm} 
}
\end{center}\vspace*{-2mm}
\caption{{\small 
The shifts $\tilde m_b\!-\!m_b$ in the quark mass \,({\bf a})\, and
$\mu_\pi^2\!-\!\tilde\mu_\pi^2$ in the kinetic operator \,({\bf b})\, 
introduced by imposing a lower cut in the photon
energy in $B\!\to\! X_s\!+\!\gamma$. Blue and maroon curves
correspond to two different ans\"{a}tze for the heavy quark distribution
function, $F_1$ and $F_2$, respectively.
}}
\end{figure}

The naive extraction of the kinetic expectation value through the
variance of the truncated distribution undercounts it even more
dramatically as Fig.~1b illustrates, since higher moments are more
sensitive to the tail of the distribution.

To gauge the sensitivity to the choice of the heavy quark
distribution function we follow Ref.~\cite{uses}\, evaluating the
effect for two ans\"{a}tze -- one exponential in $k_+$, and the other 
in $k_+^2$ yielding an even faster decreasing tail:
\beq
F_1(k_+)=N_1\,(\La\!-\!k_+)^\alpha \,e^{ck_+} \:\theta(\La\!-\!k_+),
\msp{5}
F_2(k_+)=N_2\,(\La\!-\!k_+)^\beta \,e^{-d(\La\!-\!k_+)^2}\:
\theta(\La\!-\!k_+)  \; ; 
\label{36}
\eeq
the parameters are adjusted in such a way as to yield the same actual
$m_b$ ($\La$) and $\mu_\pi^2$. (For $m_b\!=\!4.6\GeV$ and
$\mu_\pi^2\!=\!0.43\GeV^2$ we have $\alpha\!=\!2.22$ and
$\beta\!=\!0.773\,$.) Curiously, the `deficit' 
$\La\!-\!\tilde\Lambda(E_{\rm cut})$ practically does not depend on the 
choice at $E_{\rm cut}$ around $2\GeV$, and even the deficit in
$\mu_\pi^2$ is reasonably stable. 
\vspace*{.4mm}

Why are the effects of the cut so significant? They are exponential in
the inverse hadronic scale $\mu_{\rm hadr}$, but are governed by the 
{\tt hardness} ${\cal Q}\!\simeq\! m_b\!-\!2E_{\rm cut}$ rather than by $m_b$: 
\beq
\La-\tilde\Lambda(E_{\rm cut}) \;\propto\; \mu_{\rm hadr}\;
e^{-\frac{{\cal Q}}{\mu_{\rm hadr}}}\;, \msp{10}
\mu_\pi^2-\tilde \mu_\pi^2 \; \propto \;\mu_{\rm hadr}^2\;
e^{-\frac{{\cal Q}}{\mu_{\rm hadr}}}
\label{38}
\eeq
(the exponent may have a form of a power of ${\cal Q}/\mu_{\rm
  hadr}$). Even at $m_b\to \infty$ these effects survive unless
${\cal Q}$\, is made large as well!

As explained in Ref.~\cite{ckm03} the bias terms (\ref{38}) are
associated from a theoretical viewpoint with the limited (in fact,
zero) convergence radius of the OPE.  This becomes practically
relevant due to the presence of a subseries in powers of 
$1/{\cal Q}$ rather than $1/m_b$. The limitations on convergence 
appear due
to a factorial growth of the power coefficients, a rather universal
property of the OPE. In this respect one may associate this effect
with quark-hadron duality \cite{vadem}. Yet it has no features
peculiar to {\tt local} quark hadron duality violation intrinsic to
inclusive decay widths in actual Minkowski world. (For a discussion
of the notorious subtleties in the notion of quark-hadron duality,
see reviews \cite{shifioffe,vadem}). For instance, these effects are
truly exponential and do not oscillate.\footnote{Peculiarities of
real local duality would appear here only at the next-to-leading
order in $1/m_b$ and, therefore are not of much interest.}

The validity of the routinely applied expressions for the moments
with cuts for actual decays is additionally complicated by
perturbative corrections. Incorporated into fits are naive
sums of pure perturbative and pure nonperturbative terms:
\beq
M_n^{\rm np} \to M_n^{\rm np} +  M_n^{\rm pert}(\alpha_s, m_b,
E_{\rm cut})
\label{42}
\eeq
where nonperturbative corrections to the moments $M_n$ 
still do not depend on $E_{\rm cut}$. This is not 
true in general, but would hold if the actual
spectrum were exactly a convolution of the perturbative and
nonperturbative spectra,
\beq
\frac{{\rm d}\Gamma}{{\rm d}E_\gamma} = \int {\rm d}k \; \frac{{\rm 
d}\Gamma^{\rm
pert}(E_\gamma\!-\!k)}{{\rm d}E}\;
\frac{{\rm d}\Gamma^{\rm np}(\frac{m_b}{2}\!+\!k)}{{\rm d}E}
\label{52}
\eeq
{\tt provided} no cut is introduced (or its effect on the pure
nonperturbative distribution is negligible). We hasten to add,
though that the effects we consider are unrelated to this
complication and rather represent an independent phenomenon -- 
they are significant even in the complete 
absence of perturbative corrections. 

Since the perturbative effects can potentially modify the
effect, we have evaluated the cut-induced deficit in $\tilde\Lambda$ and
$\tilde\mu_\pi^2$ including short-distance corrections. Namely, we 
considered $\tilde\Lambda$, $\tilde\mu_\pi^2$ in Eqs.~(\ref{28}) for
the complete spectrum obtained by the convolution
(\ref{52}) of the perturbative and primordial (nonperturbative)
ones, and compared them to the naive sum Eq.~(\ref{42}) which would
indeed hold for a sufficiently low cut. Including the perturbative
spectrum as detailed in Ref.~\cite{uses} we found no appreciable change
in $\La\!-\!\tilde\Lambda$ or $\mu_\pi^2\!-\!\tilde\mu_\pi^2\,$ at realistic
cuts (a small increase emerged only at $E_{\rm cut}\!\lsim\! 1\GeV$).

\vspace*{2mm}
Based on these results we conclude:\vspace*{.5mm}\\
$\bullet$ the value of $m_b$ as routinely extracted from the $b\!\to\!
s+\gamma$ spectrum is to be decreased by an amount of order
$70\MeV$;\vspace*{.5mm}\\
$\bullet$ relative corrections to $\mu_\pi^2$ are even more
significant and can naturally constitute a shift of
$\,0.2\GeV^2$. This arises on top of other potential effects.
\vspace*{2mm}

\noindent
{\bf Practical implications}~~
Accepting the above shifts at face value and using the 
rather arbitrary choice $m_b\!=\!4.595\GeV$, $m_c\!=\!1.15\GeV$,
$\mu_\pi^2\!=\!0.45\GeV^2$, $\tilde\rho_D^3\!=\!0.06\GeV^3$ and 
$\rho_{LS}^3\!=\!-0.15\GeV^3$ adjusted to
accommodate $\aver{M_X^2}_{E_\ell>1\,{\rm GeV}}\,$, we
obtain\footnote{The two values for the second $E_\gamma$-moment
correspond to the two ans\" atze; they are obtained discarding higher-order
power corrections to the light-cone distribution function.}
\bea
\nonumber
\rule{0mm}{5mm}\aver{M_X^2} &\msp{-5}\simeq\msp{-5} & 
4.434\GeV^2 \msp{5}[\mbox{cf. } (4.538\pm 0.093) \GeV 
\mbox{ (New DELPHI)}\,] \\
\nonumber
\rule{0mm}{5mm}\aver{M_X^2}_{E_\ell>1.5\,{\rm GeV}} &\msp{-5}\simeq\msp{-5} & 
4.177\GeV^2 \msp{5.1}[\mbox {cf. } 4.180\GeV^2 \mbox{ (BaBar), } 
4.189\GeV^2 \mbox{ (CLEO)}\,] \\
\nonumber
\rule{0mm}{5mm}\aver{E_\ell} &\msp{-5}\simeq\msp{-5} &  1.389\GeV
\msp{6.9}[\mbox {cf. } (1.383 \pm 0.015)\GeV \mbox{ (DELPHI)}\,] \\
\nonumber
\rule{0mm}{5mm}\aver{E_\gamma}_{E_\gamma>2\,{\rm GeV}} 
&\msp{-5}\simeq\msp{-5} & 
2.329\GeV \msp{6.9}[\mbox {cf. } (2.346 \pm 0.034)\GeV  \mbox{ (CLEO)}\,]\\ 
\rule{0mm}{5mm}
\aver{E_\gamma^2\!-\!\bar E_\gamma^2}_{E_\gamma>2\,{\rm GeV}} 
&\msp{-5}\simeq\msp{-5} &\msp{-.9}
^{0.0202}_{0.0233} \,\GeV^2 \msp{5.0}
[\mbox {cf. } (0.0226 \pm 0.0066\pm0.0020)\GeV^2  \mbox{ (CLEO)}\,] \qquad
\label{56}
\eea
(experimental data are from Refs.~\cite{newdelphi,newbabar,newcleo,cleobsg}).
$E^\ell_{\rm cut}$-dependence of $\aver{M_X^2}$ is also reproduced,
see Fig.~2. The counterpart of the above exponential cut-related 
effects for the semileptonic decays has not been 
incorporated here, however.

\begin{figure}[hhh]
\begin{center}
\mbox{\psfig{file=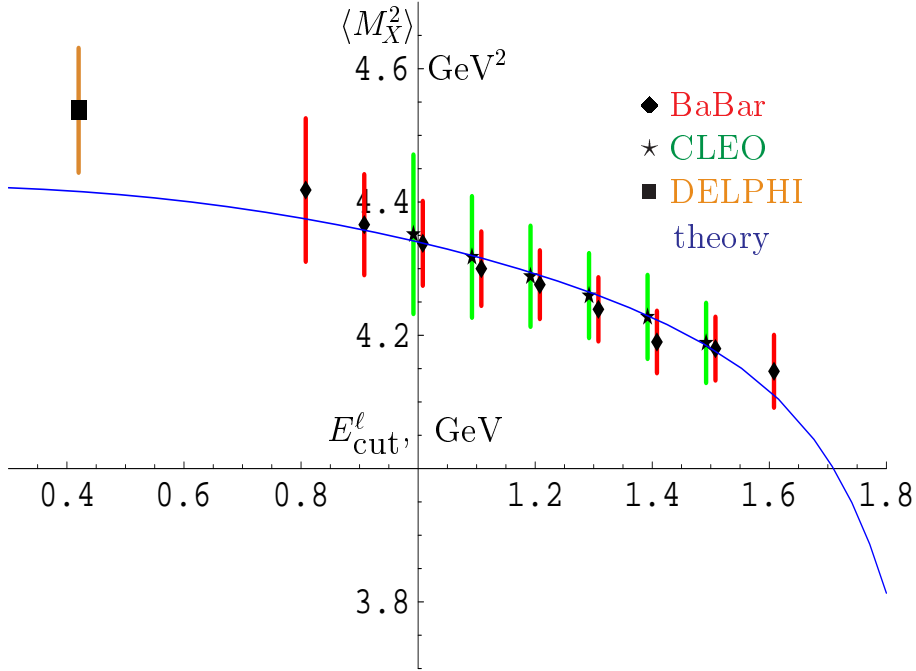,width=120mm}}
\end{center}\vspace*{-2mm}
\caption{{\small 
Experimental values of the average hadronic mass square
$\aver{M_X^2}$ in $B\!\to\! X_c\,\ell\nu$ at different lower cuts in
$E_\ell$  and  the literal OPE prediction (blue curve) for the
stated heavy quark parameters. 
The DELPHI point assumes no cut on $E_\ell$. 
}}
\end{figure}

It should be noted that the cut-induced shifts $\tilde m_b \!-\!
m_b$, $\;\tilde \mu_\pi^2 \!-\! \mu_\pi^2$ are not unambiguously
determined in terms of a few known HQ parameters, but
rather depend on the actual shape of the heavy quark distribution
function; in particular it is driven by the asymptotics 
of certain high-dimension
expectation values. The evaluation presented above relies on the most
natural assumptions about the function. Strictly speaking, they
can  vary for more contrived ans\" atze. Therefore, the estimates in 
Eqs.~(\ref{12}) can be conservatively viewed as the minimal possible
theoretical inaccuracy of the usual naive evaluations. The safest
way to tackle possible ambiguity is to resort to more inclusive
moments without too severe cuts.
\vspace*{2mm}

\noindent
{\bf Cuts in semileptonic moments.}~~
As argued in Ref.~\cite{ckm03}\, similar cut-related `exponential'
biases missed in the naive OPE applications affect the truncated
moments in the semileptonic decays as well. Their description, even 
simplified is less transparent and would be more involved,
though. In particular, the light-cone distribution functions for
$b\!\to\! s+\gamma$ is replaced by a different function, the form of which
actually is not universal. Yet the qualitative trend is expected the
same -- it should smoothly interpolate the case of literal OPE
expression at low $E_\ell$ cuts and the values at a high cut dictated
simply by the actual hadrons kinematics. This would replace step- or
$\delta$-like behavior in the formal OPE expressions. 

Numerical aspects are less certain. Keeping in mind that for
$B\!\to\! X_c\,\ell\nu $ the effective hardness 
\beq
{\cal Q}_{\rm sl}\simeq m_b-E_{\rm cut}-\sqrt{E_{\rm cut}^2+m_c^2} 
\label{80}
\eeq
at $E_{\rm cut}\!=\!1.5\GeV$ is about $1.25\GeV$ \cite{amst}, nearly 
the same as for $B\!\to\! X_s+\gamma$ with $E_\gamma \!\gsim\! 2\GeV$, we
may expect quite significant effects. To get a rough idea of the 
possible magnitude of the bias we can use a simplified rule of thumb -- assume
that $m_b$ in the semileptonic decay can be just replaced by an
effective larger value $\tilde m_b$ apparent in $b\!\to\! s+\gamma$ at
the commensurate cut yielding the same hardness. In other words,
we mimic the effect of decreasing hardness by an additional effective
nonperturbative running of the heavy quark mass at low scales. As
illustrated above, this is to increase $m_b$ (or, equivalently
decrease $\La$) by about $70\MeV$, a significant change.

Yet the semileptonic decay characteristics strongly depend on
both $m_b$ and $m_c$. To stay on the conservative side we assume
the apparent shift in $m_c$ as high as in $m_b$ (which would reflect
just the heavy flavor symmetry):
\beq
m_c \to \tilde m_c \approx m_c+(\tilde m_b\!-\!m_b)\;.
\label{90}
\eeq
In actuality the corrections to $m_c$ are typically somewhat smaller
due to $1/m_c^k$ terms.

As discussed elsewhere \cite{imprecated}, for actual $B$ decays 
both lepton moments
and $\aver{M_X^2}$ depend on more or less the same combination of
masses $m_b\!-\!0.65m_c\,$.  This means that the literal ansatz (\ref{90})
would suppress the effect by a factor of $3$ to $4$, yet $1/m_c$
corrections may be thought to softening this suppression. We then
expect the exponential terms in semileptonic decays with $E_{\rm
  cut}\!\simeq\! 1.5\GeV$ to introduce effects on the same scale as
shifting $m_b$ upward by up to $25$ to $30\MeV$ (assuming fixed
$m_c$ and other heavy quark parameters). This rule of thumb is
useful to get an idea of the ultimate theoretical accuracy one can
count on.

For example, the CLEO's cut moment 
$\,R_1\!=\!\aver{E_\ell}\raisebox{-1.3mm}{\mbox{{\tiny 
$\!E_\ell\!\! >\!\!1.5$GeV\,}}}$ is approximately given by \cite{amst}
\beq
R_1\!=\!1.776\GeV + 0.27(m_b\!-\!4.595\GeV) -
0.17(m_c\!-\!1.15\GeV) 
\msp{7}\mbox{at }
 |V_{ub}/V_{cb}|\!=\!0.08
\label{94}
\eeq
(the above mentioned values of the nonperturbative parameters are
assumed). An increase in $m_b$ by only $20 \MeV$ would then change
\beq
R_1 \to R_1+ 0.0055\GeV
\label{96}
\eeq
and would perfectly fit the central CLEO's value
$1.7810\GeV$. It is worth noting that Eqs.~(\ref{94})-(\ref{96})
make it explicit  that the imposed cut on $E_\ell$ degrades the
theoretical calculability of $R_1$ far beyond its experimental error
bars, the fact repeatedly emphasized over the last
year. Unfortunately, this was not reflected in the fits of
parameters which placed
much weight on the values of $\,R_0\,$--$\,R_2\,$ just owing to their
small experimental uncertainties, whilst paying less attention to actual
theoretical errors.\footnote{N.U.\ acknowledges that a similar in spirit
criticism of the theory error treatment in \cite{gfit} 
was expressed by D.\,Hitlin at the BaBar Workshop, SLAC December 2002.}

Similar reservations apply to the quality of theory for CLEO's $\,R_2$
representing the second moment with the cut, with the only
difference that the effective hardness only deteriorates for higher
moments. 

The ratio $R_0$ is the normalized decay rate with the cut on
$E_\ell$ as high as $1.7\GeV$, and for it hardness ${\cal Q}$ is
below $1\GeV$. A precision -- beyond just semiquantitative -- treatment
of nonperturbative effects is then questionable, and far more significant
corrections should be allowed for. 
\vspace*{2mm}

We note that there is a good agreement of most data referring to
sufficiently `hard' decay distributions with the theory based on the
OPE in QCD, in the `robust' OPE approach. The latter was put forward
\cite{amst} to get rid of vulnerable and unnecessary assumptions of
the usually employed fits of the data.
This consistency likewise refers to the absolute values of the 
heavy quark parameters
necessary to accommodate the data. Theory anticipates, however that
the expansion becomes deceptive with increase of the experimental
cuts. Here we have addressed the most obvious effects, those from the
variety of `exponential' terms in the effective hardness. While
presently not amenable to precise theoretical treatment, they can
be estimated using quite natural assumptions and are found to be far
too significant for $E^{\ell}_{\rm cut}\!\gsim\! 1.5\GeV$ 
and $E^{\gamma}_{\rm cut}\!\gsim\! 2\GeV$ often employed in
experiment. Taking these corrections at face value and incorporating
in our predictions, we get a good, more than qualitative agreement
with ``less short-distance'' inclusive decays as well. 
\vspace*{2.5mm}

{\tt To summarize},
the cuts essentially decreasing the hardness in
$B$ decays introduce terms some of which are exponentially 
suppressed though only in the effective hardness, but not suppressed by 
powers of the heavy quark
mass. They significantly change the extracted values of the heavy quark
parameters leading to an apparent suppression of the magnitude of the 
nonperturbative ones in $B$ mesons. Accounting for such effects appears 
necessary in $B\!\to\!X_s+\gamma$ decays unless the cut on
photon energy is pushed well below present $2\GeV$. This brings the
practical predictions for various inclusive characteristics into a
good agreement. Moreover, it seems that the second moment of the 
photon spectrum discarded for the fits so far, does impose
informative constraints provided these effects are properly incorporated.

\vspace*{1mm}

{\bf Acknowledgments:} N.U. is indebted to P.~Gambino for joint work
heavily used in the present analysis. He is also thankful to
O.~Buchmueller and U.~Langenegger for many discussions which
initiated this study, and for useful comments. 
This work was supported in part by the NSF under grant number PHY-0087419.

\end{document}